\documentclass[aps,a4paper,twocolumn,pre]{revtex4}
\pdfoutput=1
\usepackage{graphicx}
\usepackage{amssymb}
\usepackage{epstopdf}
\usepackage{natbib}

\begin{document}
\title{Effect of excluded volume interactions on the interfacial properties of colloid-polymer mixtures}
\author{Andrea Fortini$^{1}$\footnote{Present address: Department of Physics, Yeshiva University, 500 West 185th Street, New York, NY 10033, USA}}
\author{Peter G. Bolhuis$^{2}$} 
\author{Marjolein Dijkstra$^{1}$}
\affiliation{{\rm(1)} Soft Condensed Matter, Debye Institute for NanoMaterials science,Utrecht University,
   Princetonplein 5, 3584 CC Utrecht, The Netherlands. \\
{\rm(2)} van't Hoff Institute for Molecular Sciences, University of
Amsterdam, Nieuwe Achtergracht 166, 1018 WV Amsterdam, The
Netherlands.}

\begin{abstract}
We report a numerical study of equilibrium phase-diagrams and interfacial properties of bulk and confined colloid-polymer mixtures using grand canonical Monte Carlo simulations.
Colloidal particles are treated as hard spheres, while the polymer chains are described as soft
repulsive spheres. The polymer-polymer, colloid-polymer, and
wall-polymer interactions are described by density-dependent
potentials derived by Bolhuis and Louis [Macromolecules, 35
(2002), p.1860]. We compared our results with those of the
Asakura-Oosawa-Vrij model, that treats the polymers as ideal
particles. We find that the number of polymers needed to drive the
demixing transition is larger for the interacting polymers, and
that the gas-liquid interfacial tension is smaller. When the
system is confined between two parallel hard plates, we find
capillary condensation. 
Compared with the AOV model, we find that the excluded volume interactions between the polymers suppress capillary condensation. 
In order to induce capillary condensation, smaller undersaturations and smaller plate separations are needed in comparison 
with ideal polymers.
\end{abstract}
\maketitle

\section{Introduction}

\begin{figure}[htbp]
   \centering
   \includegraphics[width=8cm]{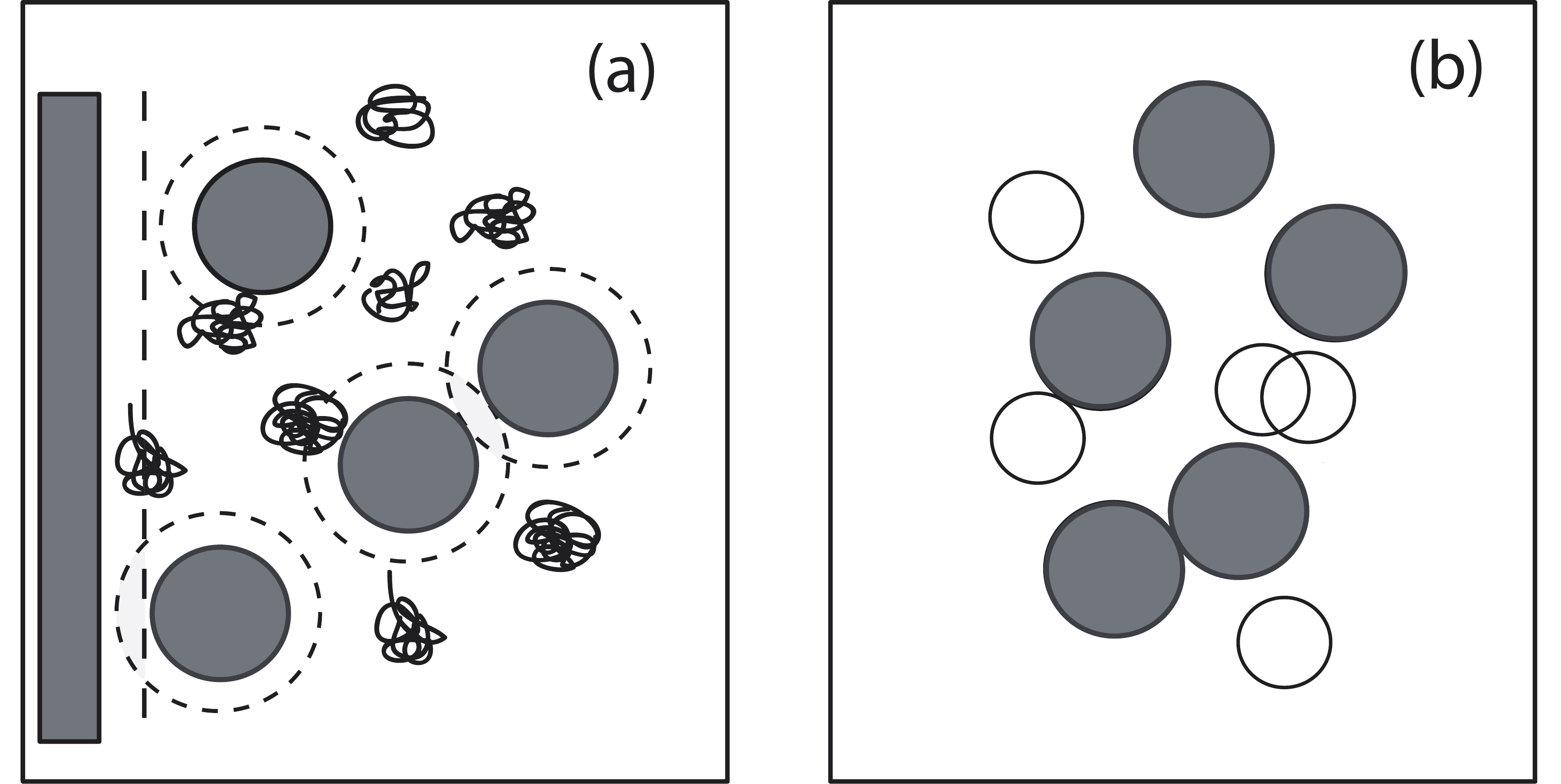}
   \caption{(a) Illustration of a mixture of colloids and polymer
   chains in contact with a hard wall. Depletion zones (dashed lines)
   and overlap zones (light grey) are also depicted. (b) Illustration
   of a model colloid-polymer mixture in the polymer-as-soft-sphere approach.}
   \label{fig:model}
\end{figure}

Mixtures of colloids and
polymers~\cite{Poon2002,Tuinier2003,Brader2003} are simple model
systems that have been studied extensively in the past years.
Provided the size and the number of polymers are sufficiently
high, such mixtures can phase-separate into a \emph{colloidal gas}
phase that is poor in colloids and rich in polymers, and a
\emph{colloidal liquid} phase that is rich in colloids and poor in
polymers. While it is well established that the interactions between sterically stabilized
colloidal particles are well described by the hard sphere
potential,~\cite{Pusey1986} the interactions between the polymers
in the mixture can be in general very complicated. Nevertheless,
if we consider the case of flexible polymer chains in a ``good
solvent'' conditions, we can assume that  the excluded volume
interactions between the chains are small and that the chains
cannot penetrate the colloidal particles. The mechanism behind the
demixing transition is then easily explained. In Fig.
\ref{fig:model}(a), we illustrate the mixture of spherical
colloids and polymer chains. Around each colloid there is a
depletion region prohibited to the polymers due to the hard-core
interactions. If two colloids approach each other, so that two
depletion zones overlap there is an increase in free volume for
the polymer chains, i.e. an increase in entropy. The increase in
entropy can be described by an effective attractive interaction
between the colloidal particles. A similar depletion mechanism
occurs between a hard wall and the colloidal particles. If the
polymers do not adsorb at the wall they are excluded from a region
close to the wall. The overlap between the depletion zone at the
wall and the depletion zone of one colloid induces an increase in
free volume and, hence, in entropy.

One particular simple model for colloid-polymer mixtures is the
Asakura-Oosawa-Vrij (AOV)
model~\cite{Asakura1954,Asakura1958,Vrij1976} that describes the
polymer chains as spherical particles with a radius equal to the
radius of gyration of the polymer. Furthermore, polymer spheres
can freely overlap, while they are excluded from a centre of mass
distance from the colloidal particles. The AOV model has been
studied extensively in the past years, and it was shown that it
describes qualitatively the bulk~\cite{Gast1983,Lekkerkerker1992,Meijer1994,
Dijkstra1999a,Hoog1999,Chen2000,Schmidt2000,Schmidt2002,Vink2004}
and interfacial phase
behavior~\cite{Brader2001,Brader2002,Dijkstra2002,Wijting2003,Aarts2003,Wijting2003a}
of mixtures of colloids and polymers. A similar level of agreement
was found for the interfacial tension of the
gas-liquid~\cite{Vink2004,Aarts2005,Fortini2005a} and wall-fluid
interfaces,~\cite{Wessels2004a,PPFW04a,Fortini2005a} and for the
phase behavior of confined
systems.~\cite{Schmidt2003,Lee2003,Schmidt2004b,Aarts2004b,Fortini2006a,Vink2006,Vink2006a,Virgiliis2007}

The quantitative discrepancies between experimental results and
the AOV model results  can arise from a number of reasons, like
non-ideal solvent conditions,~\cite{Schmidt2002b} colloid-induced
polymer compression,~\cite{Denton2002} effect of charges on the
colloidal surface,~\cite{Denton2005,Fortini2005} or  polymer
excluded volume interactions. In this article, we will concentrate
on the latter aspect. The simplest inclusion of polymer
interactions was done by introducing a step function interaction
between the polymers, i.e. an energy penalty for the overlaps of
two polymers. The step potential was used to study the bulk phase
diagram and interfacial
tension~\cite{Schmidt2003a,Wessels2004a,Vink2005a} as well as the
stability of the floating liquid phase in sedimenting
colloid-polymer mixtures~\cite{Schmidt2004} with a geometry-based
density functional theory (DFT).  This approach gives results that
are in better agreement with experiments when compared against the
AOV model, but the height of the step potential must be introduced
as an additional free parameter. Furthermore, we expect  the
polymer-colloid interaction to be modified as well when
considering excluded volume interactions between polymers. Other
theoretical approaches have been developed to study the effect of
polymer interactions in colloid-polymer mixtures.
\citet{Aarts2002} extended the free volume
theory~\cite{Lekkerkerker1992} to include polymer interactions,
and studied  the gas-liquid interfacial tension with the square
gradient approximation
approach.~\cite{Brader2000,Aarts2002,Moncho-Jorda2003,Aarts2004a}
A one-component perturbative DFT that includes excluded volume interactions that uses the approach
of Ref.~\cite{Louis2002} was developed by~\citet{Moncho-Jorda2005}
to study confined systems and the gas-liquid tension.

Another approach is to describe the polymers as  soft
spheres,~\cite{Louis2000,Bolhuis2001,Bolhuis2002a} with  effective interactions derived
from inversion of the centre of mass (CM) correlation functions in
lattice Self-Avoiding-Walk (SAW) polymer simulations. This
approach generates soft density-dependent potentials for the
polymer-polymer and colloid-polymer interactions, that give
accurate simulation results for the bulk phase behavior,~\cite{Bolhuis2002} that are in quantitative agreement with
experimental results.
In a similar approach, proposed
by~\citet{Jusufi2001}, the effective potentials are derived from
off-lattice molecular dynamics simulations of SAW polymer chains.
These potentials were used in Monte Carlo simulations to study the
bulk phase behavior~\cite{Dzubiella2002,Rotenberg2004} and the
gas-liquid interfacial tension~\cite{Vink2005} of colloid-polymer
mixtures, leading to results in quantitative agreement with
experiments.

In this work, we study the effect of excluded volume interactions
on the interfacial properties and phase behavior of confined
colloid-polymer mixtures with Monte Carlo simulations. We simulate
a binary mixture with the density-dependent potentials derived
in Ref.~\cite{Bolhuis2002a}, and compare the simulation results with
those for the AOV model.

\section{Model}
\label{ts:mod}
The colloids are treated as hard spheres and the corresponding pair potential reads
\begin{equation}
\beta v_{\text{cc}}(R_{ij})= \left \{ \begin{array}{ll}
\infty & \textrm{ if $R_{ij} < \sigma_{\text{c}}$ } \\
0 & \textrm{otherwise},
\end{array} \right.
\end{equation}
where $R_{ij}=|\vec{R}_i-\vec{R}_j|$ is the distance between two
colloidal particles, with $\vec{R}_i$ the position of the centre
of mass of colloid $i$, $\beta \equiv 1/k_B T$, $k_B$
Boltzmann's constant, and $T$ the temperature. For the coarse-grained effective
polymer-polymer, colloid-polymer, and wall-polymer potentials we
use the expressions of Ref.~\cite{Bolhuis2002a} obtained from
microscopic simulations of SAW polymer chains consisting of 500
segments on a lattice and a radius of gyration $R_{g}$=16.83
lattice units at zero concentration. We introduce the overlap
concentration $\rho^{*}$ defined by the equation $4/3 \pi \rho^{*}
R_{g}^{3}=1$, and the polymer reservoir packing fraction
$\eta_{p}^{r}=\rho_{\rm p}^{r}/\rho^{*}$, with $\rho_{\rm p}^{r}$
the density in the reservoir of pure polymers in osmotic
equilibrium with the two-component system of interest.

The effective density-dependent polymer-polymer interactions,~\cite{Bolhuis2002a} read
\begin{equation}
\beta v_{\rm pp}(r_{ij},\eta_{p}^{r})=\sum_{k=1}^{3}
a_{k}(\eta_{p}^{r}) \exp[(-r_{ij}/(R_{g} b_{k}(\eta_{p}^{r})))^{2}],
\label{E:vpp}
\end{equation}
where $R_{g}$ is the radius of gyration, $r_{ij}=|\vec{r_i}-\vec{r_j}|$ is the distance between two
polymers, with $\vec{r_i}$ the position of the centre of mass of
polymer $i$. The density-dependent parameters are linear in the
density $a_{k}=a_{k}^{0}+a_{k}^{1} \eta_{p}^{r}$, and
$b_{k}=b_{k}^{0}+b_{k}^{1} \eta_{p}^{r}$. All coefficients, except $b_{k}^{3}$, are given in Table \ref{tab:pp}. 
The coefficients $b_{k}^{3}$ are fixed by imposing the equality of
the mean field equation of state
\begin{equation}
\beta P/\rho_{p}=1+\rho_{p} \beta \hat v(0; \rho_{p})/2 \ ,
\label{te:EOS}
\end{equation}
for the fitted potentials and the SAW simulations, where the function
\begin{equation}
\beta \hat v(0; \rho_{p})=4 \pi \int r^{2} \beta v_{\rm
pp}(r,\rho_{p}) dr
\end{equation}
is the $k=0$ component of the Fourier transform of the polymer-polymer pair potential.
In practice, the condition is satisfied by imposing the equality between
\begin{equation}
\beta \hat v(0;\eta_{p}^{r}) = \pi^{3/2} \sum_{i=1}^{3}
a_{i}(\eta_{p}^{r})b_{i}(\eta_{p}^{r})^{3} \ ,
\end{equation}
derived from Eq. (\ref{E:vpp}) and
\begin{eqnarray}
\lefteqn{\beta \hat v(0;\eta_{p}^{r}) = 4 \pi( 1.2902 +}\nonumber \\
&&0.28132
~\eta_{p}^{r}+0.13676
~(\eta_{p}^{r})^{2}-0.040892~(\eta_{p}^{r})^{3}) \label{te:vhat}
\end{eqnarray}
derived using the potentials obtained from the inversion of the
radial distribution function determined from SAW simulations.

\begin{table}[htdp]
\caption{Coefficients for the density-dependent parameters of the
polymer-polymer interaction potential $v_{\rm pp}$ defined in
Eq.~(\ref{E:vpp}).}
\begin{center}
\begin{tabular}{crrr}
\hline
 & k=1 & k=2 &k=3 \\
\hline
$a_{k}^{0}$ & 1.47409 & -0.23210 & 0.63897  \\
$a_{k}^{1}$ & -0.07689 & 0.03132 &  0.24193   \\
\hline
$b_{k}^{0}$ & 0.98137 & 0.42123    & -\\
$b_{k}^{1}$ & -0.05681 & -0.02628 & -\\
\hline
\end{tabular}
\end{center}
\label{tab:pp}
\end{table}

The concentration-dependent colloid-polymer potential
reads~\cite{Bolhuis2002}
\begin{eqnarray}
\lefteqn{\beta v_{\rm cp}(|\vec{R}_i-\vec{r}_j|,\eta_{p}^{r})
= \sum_{k=1}^{2} c_{k}(\eta_{p}^{r})
 \times }\nonumber \\
&&\exp[-((|\vec{R}_i-\vec{r}_j|-e_{k}(\eta_{p}^{r}))/(R_{g} d_{k}(\eta_{p}^{r})))^{2}],
\label{E:vcp}
\end{eqnarray}
where $|\vec{R}_i-\vec{r}_j|$ is the distance between colloid $i$
and polymer $j$. The density-dependent parameters are linear in
density, i.e., $c_{k}=c_{k}^{0}+c_{k}^{1} \eta_{p}^{r}$,
$d_{k}=d_{k}^{0}+d_{k}^{1} \eta_{p}^{r}$, and
$e_{k}=e_{k}^{0}+e_{k}^{1} \eta_{p}^{r}$. The coefficients are
given in Table \ref{tab:cp} for size ratio q=1.05.

\begin{table}[htdp]
\caption{Coefficients for the density-dependent parameters of the
colloid-polymer interaction potential $v_{\rm cp}$ defined in
Eq.~(\ref{E:vcp}) for size ratio q=1.05.}
\begin{center}
\begin{tabular}{crr}
\hline
 & k=1 & k=2  \\
\hline
$c_{k}^{0}$ & 5.5610 & 1.8477   \\
$c_{k}^{1}$ & -0.8042 & 1.4759   \\
\hline
$d_{k}^{0}$ & 0.7751 & 1.2720   \\
$d_{k}^{1}$ & -0.1151 & 0.1052   \\
\hline
$e_{k}^{0}$ & 0.4082 & 0.0   \\
$e_{k}^{1}$ & 0.1410 & 0.0   \\
\hline
\end{tabular}
\end{center}
\label{tab:cp}
\end{table}%

\begin{figure}[htbp]
   \centering
   \includegraphics[width=8cm]{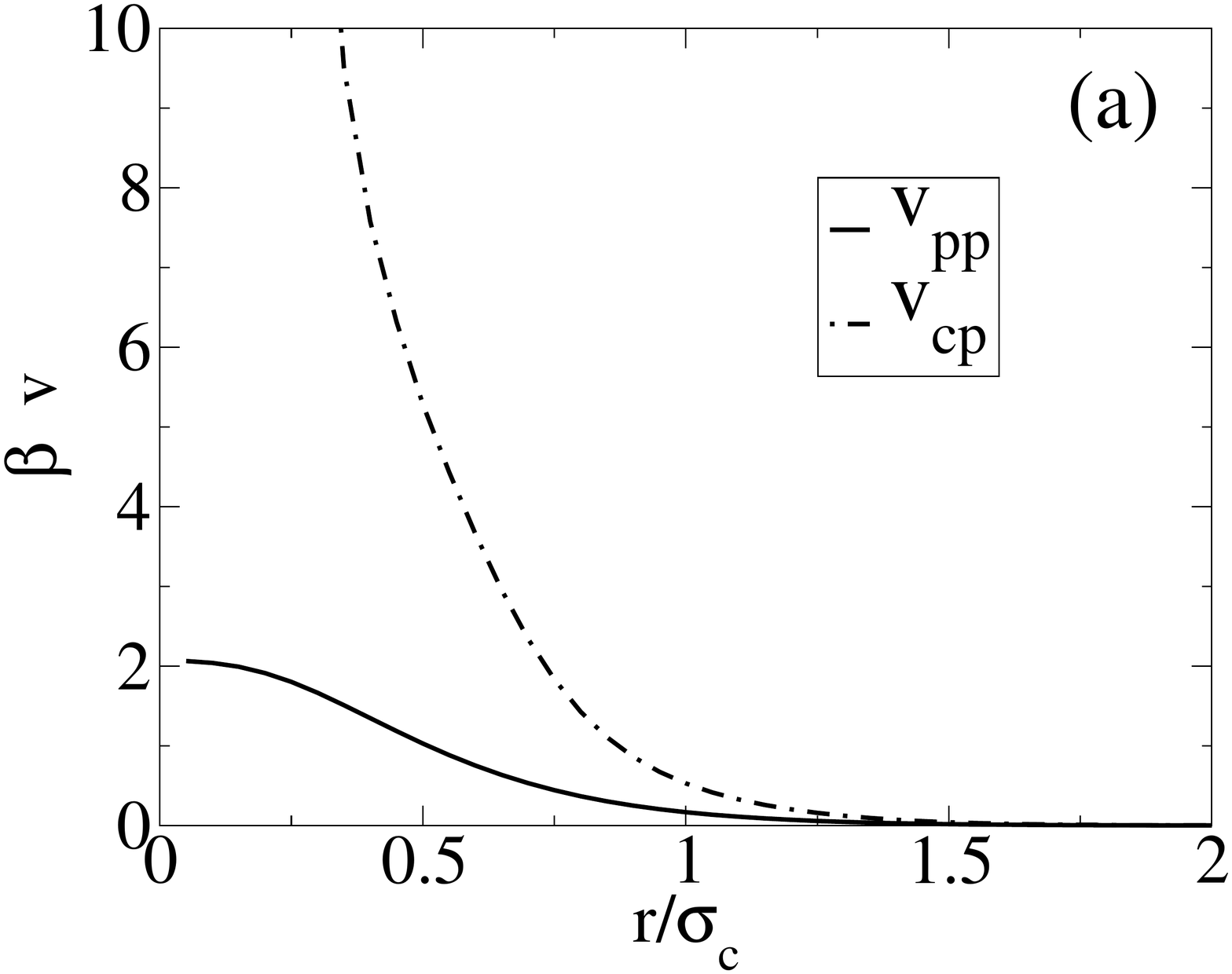}
      \includegraphics[width=8cm]{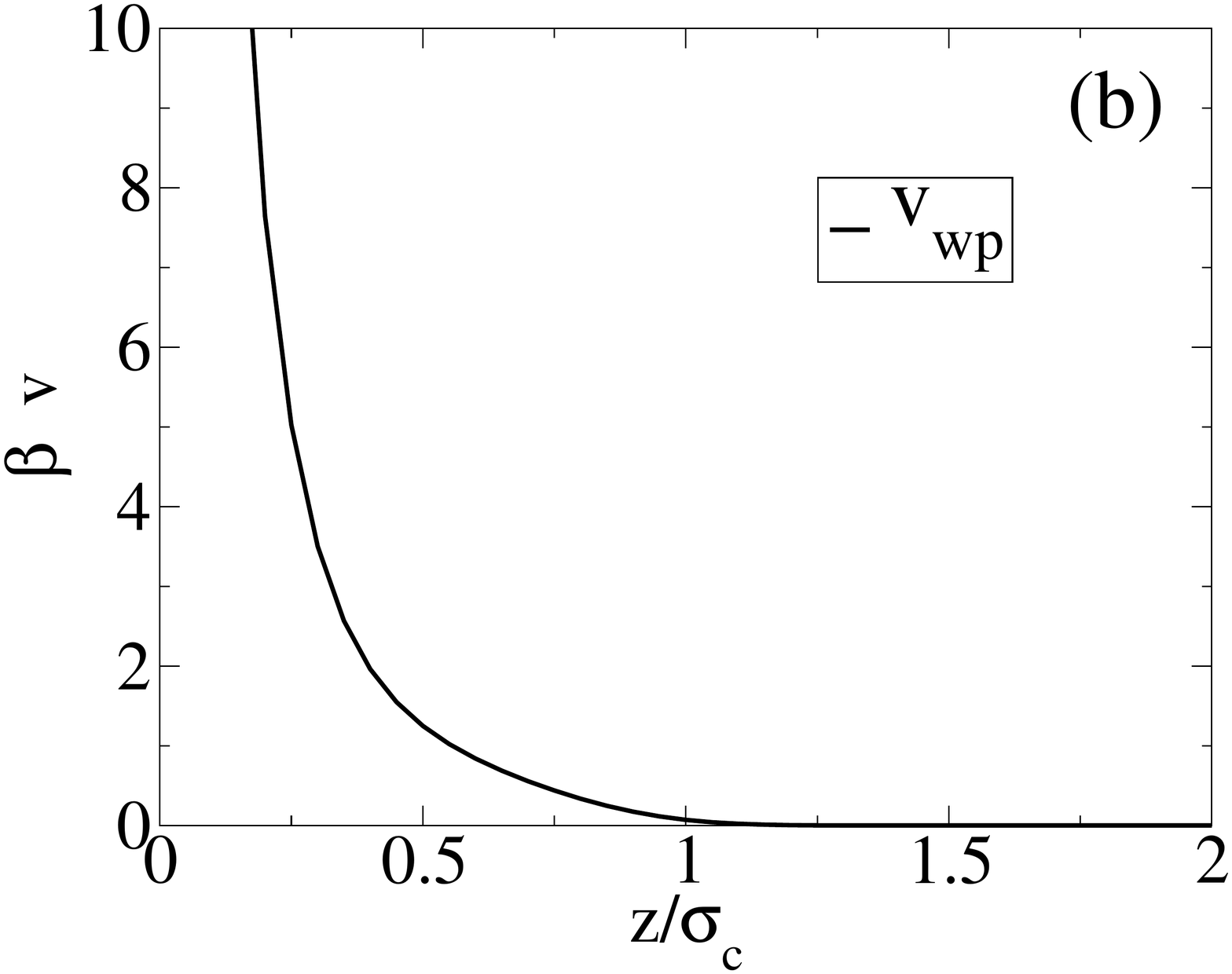}
   \caption{Interaction potentials for polymer density $\eta_{p}^{r}=1.02995$. (a) Pair potential between
   polymers $\beta v_{\rm pp}$ (full line), and between polymers and colloids $\beta v_{\rm cp}$ (dashed line). (b) Wall-polymer potential $\beta v_{\rm wp}$ as a function of the distance $z/\sigma_{c}$ of the polymer centre of mass from the wall.}
   \label{fig:pot}
\end{figure}

The interaction between colloidal particles and the hard wall is
hard-sphere-like, that is the colloidal particles cannot penetrate
the walls. The interaction between polymers and the hard
wall~\cite{Bolhuis2002a}  reads
\begin{eqnarray}
\lefteqn{ \beta v_{\rm wp}(z,\eta_{p}^{r})=f_{0}(\eta_{p}^{r}) \times} \nonumber\\
& &\exp[f_{1}(\eta_{p}^{r}) z/R_{g}+ f_{2}(\eta_{p}^{r})z^{2}/R^{2}_{g}
+f_{3}(\eta_{p}^{r})z^{3}/R^{3}_{g}],
\label{E:vwp}
\end{eqnarray}
where $z$ is the distance between the wall and the centre of mass
of the polymer. The parameters have a quadratic density-dependence
$f_{k}(\eta_{p}^{r})=f_{k}^{0}+f_{k}^{1}
\eta_{p}^{r}+f_{k}^{2}(\eta_{p}^{r})^{2}$, with $k$=0, 1, 2, 3.
The coefficients are given in Table \ref{tab:wp}. In Fig.
\ref{fig:pot}, we show the effective polymer-polymer,
colloid-polymer, and wall-polymer interactions for
$\eta_{p}^{r}=1.02995$ as an example.

\begin{table}[htdp]
\caption{Coefficients for the density-dependent parameters of  the
wall-polymer interaction potential $v_{\rm wp}$ defined in
Eq.~(\ref{E:vwp}).}
\begin{center}
\begin{tabular}{crrrr}
\hline
 &k=0&  k=1 & k=2 &k=3 \\
\hline
$f_{k}^{0}$ & 62.7242  & -6.4093 & 2.5081 & -0.6904  \\
$f_{k}^{1}$ & 56.4595 & -3.8880 & 5.1562 & -1.5519\\
$f_{k}^{2}$ & -29.9283 & 2.0442& -2.1336 & 0.5973 \\
\hline
\end{tabular}
\end{center}
\label{tab:wp}
\end{table}%

A final note on the potentials described in this section is in order.
Due to a small error in the calculation of the radius of gyration in Ref.~\cite{Bolhuis2002a}, all the equations and coefficients presented here were parameterized assuming a radius of gyration $R_{g}$=16.495 lattice units, instead of the correct value of $R_{g}$=16.83. The simulations we carried out for this work, as well as the simulations presented on Ref.~\cite{Bolhuis2002}, were done imposing a size ratio of $q=1.03$ (corresponding to the radius of gyration $R_{g}$=16.495) in the calculations. All the results have been interpreted  using the correct value for the size ratio $q=1.05$ (corresponding to the radius of gyration $R_{g}$=16.83) . 
This correction only change the value of the polymer packing fractions $\eta_{p}$ and $\eta_{p}^{r}$, while the colloid packing fraction $\eta_{c}$ is unaffected.

\section{Method}
We carried out Monte Carlo simulations in the grand canonical
ensemble, i.e. with fixed volume, temperature, and chemical
potentials $\mu_c$ and $\mu_p$ of colloids and polymers,
respectively. For each value of $\mu_p$ we determined the
potentials given in the previous section, at the polymer reservoir
packing fraction $\eta_{p}^{r}(\mu_{p})$ calculated by inversion
of the equation of state
\begin{eqnarray}
\lefteqn{\frac{\mu_{p}}{k_{B}T} R_{g}^{3}= \log(\rho_{p}^{r} R_{g}^{3})+ 0.04658 +} \nonumber \\
&&  11.05 \rho_{p}^{r} R_{g}^{3} +  35.48 (\rho_{p}^{r} R_{g}^{3})^{2} -15.71 (\rho_{p}^{r} R_{g}^{3})^{3} \ .
\label{EQ:eos1}
\end{eqnarray}
This equation was derived by integrating the Gibbs-Duhem equation
with the pressure given by Eq. (\ref{te:EOS}) and Eq.
(\ref{te:vhat}).

\begin{figure}[htbp]
   \centering
   \includegraphics[width=8cm]{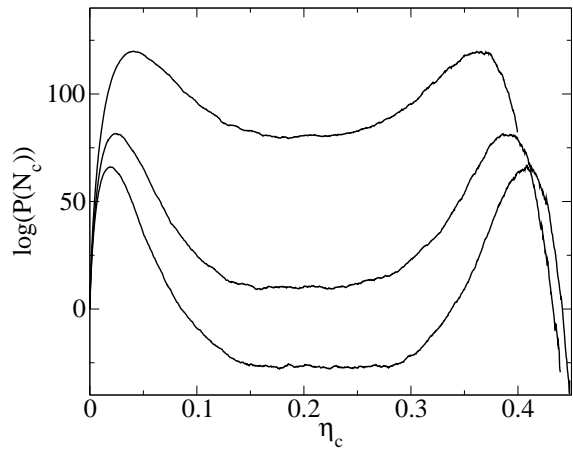}
   \caption{Logarithm of the probability $P(N_{c})$ (not normalized) as a function of the colloid packing fraction $\eta_{c}$ for a simulation box with dimensions 12x12x16 $\sigma^{3}$ at varying polymer reservoir packing fraction $\eta_{p}^{r}$=1.14, 1.20, and 1.23, from top to bottom. All state points are at coexistence.}
   \label{fig:prob}
\end{figure}

To study phase coexistence, we sample the probability
$P(N_c)|_{z_c,\eta_p^r}$ of observing $N_c$ colloids in a volume
$V$ at fixed colloid fugacity $z_c$ and fixed polymer reservoir
packing fraction $\eta_p^r$, using the successive umbrella
sampling \cite{Virnau2004}. We use the histogram reweighting
technique to obtain the probability distribution for any $z_c'$
once $P(N_c)|_{z_c,\eta_p^r}$ is known for a given $z_c$:
\begin{equation}
\ln P(N_c)|_{z_c',\eta_p^r} = \ln P(N_c)|_{z_c,\eta_p^r} +
N_c \ln\left(\frac{z_c'}{z_c}\right) \label{histogram}.
\end{equation}
At phase coexistence, the distribution function $P(N_c)$ becomes
bimodal with two separate peaks of equal area for the colloidal
liquid and gas phases.  We
determine which $z_c'$ satisfies the equal area rule
\begin{equation}
\int_{0}^{\langle N_c \rangle }P(N_c)|_{z_c',\eta_p^r}dN_c =
\int_{\langle N_c \rangle }^{\infty} P(N_c)|_{z_c',\eta_p^r}dN_c,
\end{equation}
with the average number of colloids
\begin{equation}
\langle N_c \rangle=
\int_{0}^{\infty}N_cP(N_c)|_{z_c',\eta_p^r}dN_c,
\end{equation}
using the histogram reweighting equation (\ref{histogram}). The
simulations are carried out in a rectangular box $V = L \times L
\times H$, and the sampling of the probability ratio $P(N)/P(N+1)$
is done, in each window, until the difference between two
successive samplings of the probability ratio is smaller than $5
\times 10^{-4}$. An example of the sampled probability distributions is given in Fig.~\ref{fig:prob}. 

We used single particle insertion/deletion 
of colloids and polymers. The typical 
acceptance probabilities for the insertion/deletion of colloids were
between 4\% to 1\% (from low to high colloid density) 
for state points close to the critical points and from 0.1\% to 0.01\% at high $eta_p^r$. 
On the other hand, the acceptance probabilities for the insertion/deletion of polymers were always larger then 40\%.
The low insertion/deletion probabilities of colloidal particles have only an effect on the efficiency of the algorithm. 
As shown by the smooth probability distributions in Fig.~\ref{fig:prob} our simulations were long enough to get good data. 

The liquid-gas interfacial tension $\gamma_{\lg}$
 is obtained from $P(N_c)|_{z_c',\eta_p^r}$ at
coexistence~\cite{Binder1982}
\begin{equation}
\gamma_{\lg} =\frac{1}{2 L^2}\left\lbrack
\ln \left(\frac{P(N_{c,\text{max}}^g )+P(N_{c,\text{max}}^l)}{2}\right) -
\ln(P(N_{c,\text{min}}))\right \rbrack
\end{equation}
where $P(N_{c,\text{max}}^g)$ and $P(N_{c,\text{max}}^l)$ are the
maxima of the gas and liquid peaks, respectively, and
$P(N_{c,\text{min}})$ is the minimum between the two peaks.

\section{Results}
\begin{figure}[htbp]
   \centering
   \includegraphics[width=8cm]{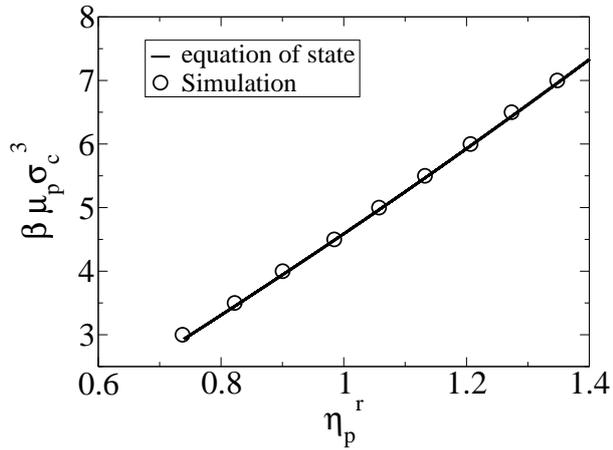}
   \caption{The chemical potential $\beta \mu_{p} \sigma_{c}^{3}$ as a
   function of the polymer packing fraction $\eta_{p}^{r}$ in a
   system of pure polymers. The equation of state (\ref{EQ:eos1})
   (solid line) is compared with the results of grand canonical
   Monte Carlo simulations (circles) of pure polymers interacting
   with the interaction potential  (\ref{E:vpp}).}
   \label{fig:eos}
\end{figure}

In Sec.~\ref{ts:mod} we explained the straightforward, but
nontrivial procedure for generating the interaction potentials. It
is therefore important to check the internal consistency of our
calculations. Fig.~\ref{fig:eos} shows the predictions of the
equation of state (\ref{EQ:eos1}) plotted against simulations
results of a grand canonical simulation of pure polymers
interacting with the potential (\ref{E:vpp}). In the range of
chemical potentials that are relevant for the gas-liquid
separation the simulation results are consistent with the equation
of state.

\subsection{Bulk phase behavior and gas-liquid interfacial tension}
\begin{figure}[htbp]
   \centering
   \includegraphics[width=8cm]{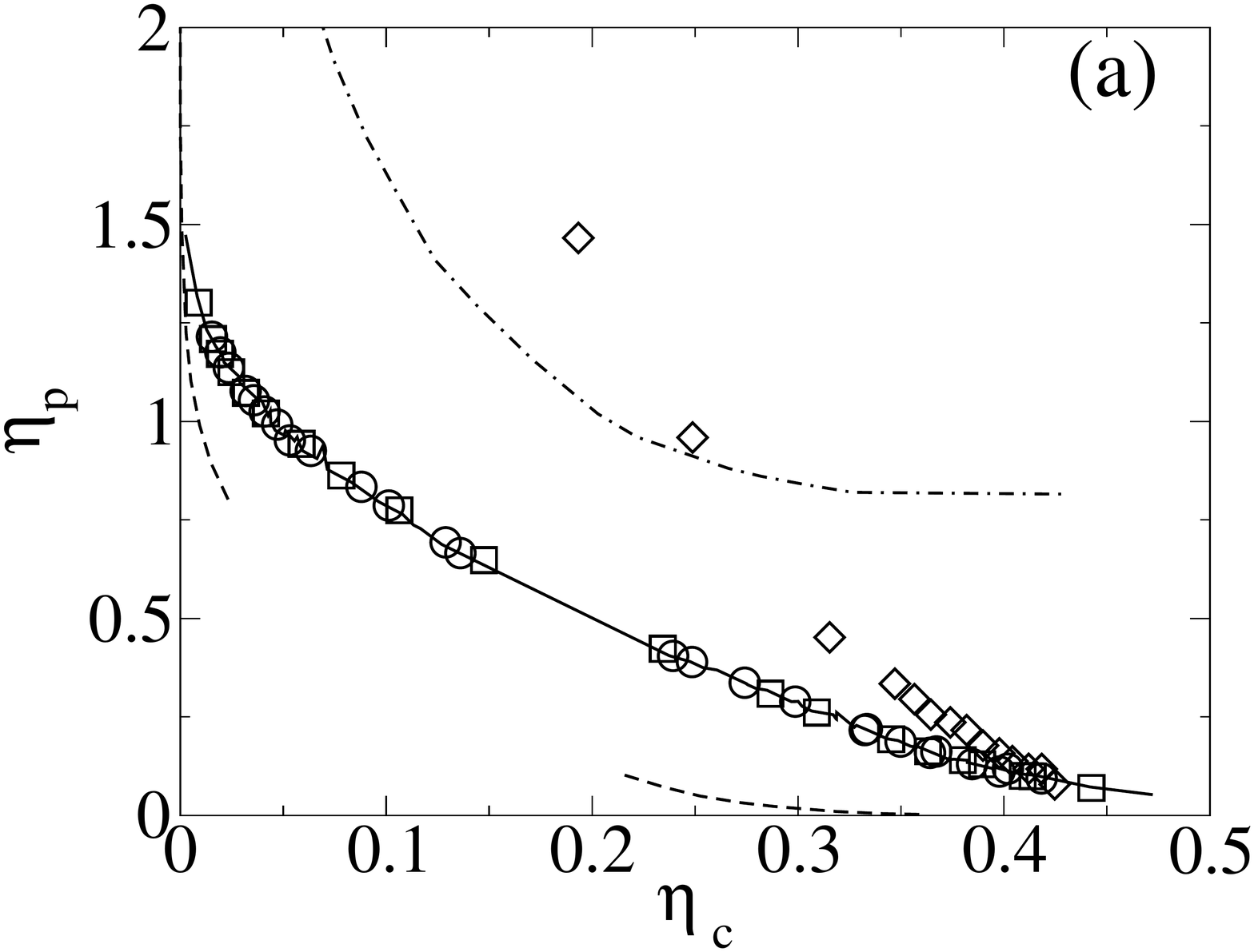}
      \includegraphics[width=8cm]{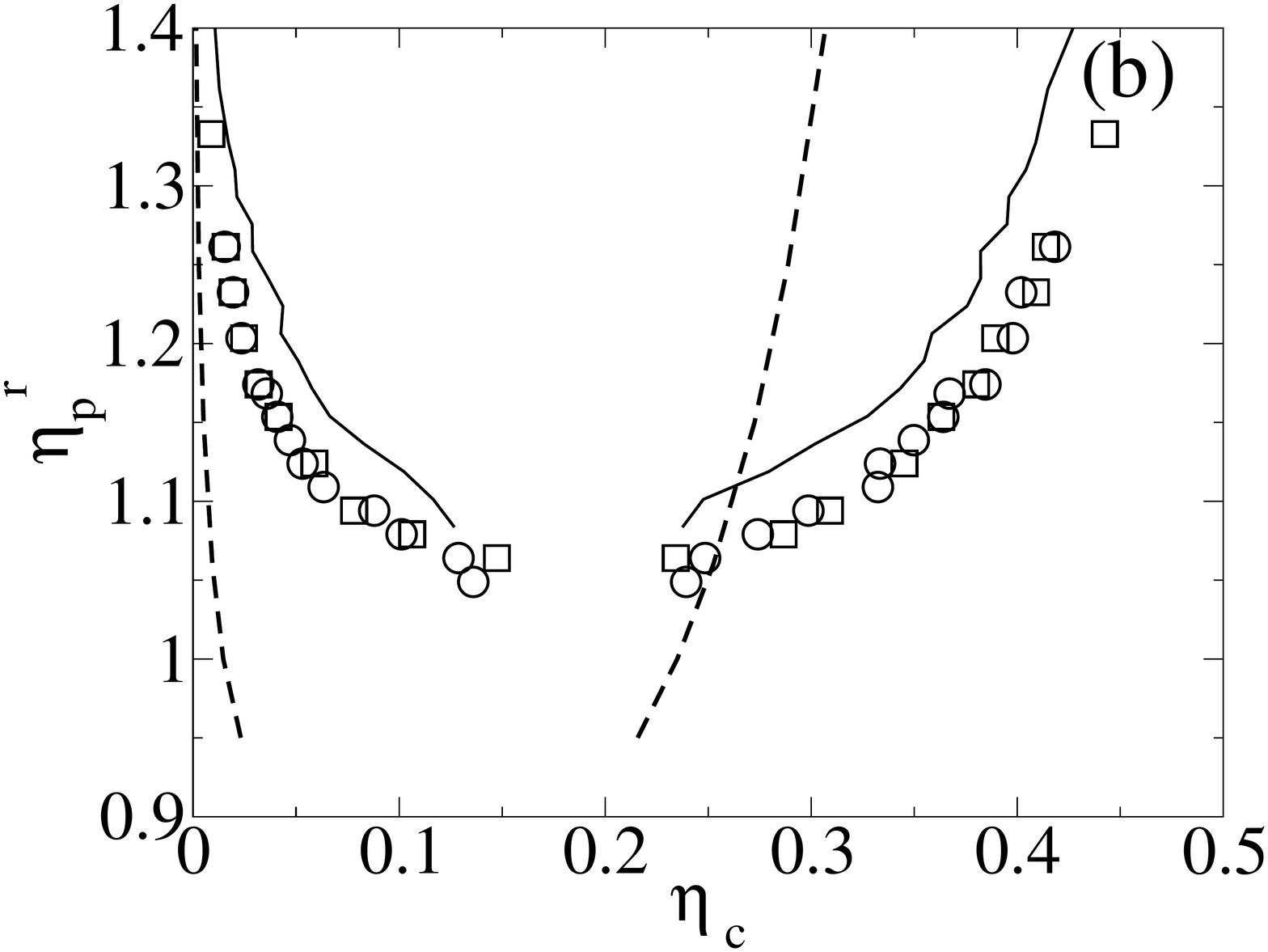}
   \caption{Bulk phase diagram of colloid-polymer mixtures with
   a size ratio $q$=1.05 for volume V=516 $\sigma_{c}^{3}$ (circles)
   and V=2304 $\sigma_{c}^{3}$ (squares). Also shown are the binodals
   of the AOV model with size ratio $q=1.0$ (dashed lines), and the results of Ref.~\cite{Bolhuis2002} (solid line). (a) Polymer
   packing fraction $\eta_p$, colloid packing fraction $\eta_c$ representation.
   Shown are also the results of the free volume theory with polymer interactions~\cite{Aarts2002}
   (dotted-dashed line) and the experimental results of~\citet{Hoog1999} (diamonds).
   (b) Polymer reservoir packing fraction $\eta_p^r$, colloid packing fraction $\eta_c$ representation. }
   \label{fig:bulk}
\end{figure}

In Fig.~\ref{fig:bulk} we present the bulk phase diagram obtained
from grand canonical Monte Carlo simulations using successive
umbrella sampling and histogram reweighting. In particular,
Fig.~\ref{fig:bulk}(a) shows the phase diagram in the polymer
packing fraction $\eta_{p}$, colloid packing fraction $\eta_{c}$
representation. These results are consistent with the findings
of Ref.~\cite{Bolhuis2002}.
The free volume theory~\cite{Lekkerkerker1990}  extended to
include excluded volume polymer interactions,~\cite{Aarts2002}
overestimates the simulation results by almost a factor of two.
This result may be explained by the renormalisation group theory
expression used in Ref.~\cite{Aarts2002} to evaluate the polymer
interactions, that underestimates the correlation length of the
polymers. Since the polymers are effectively smaller, a higher
number of polymers is needed to drive the demixing transition.
Also shown are the experimental results of~\citet{Hoog1999}. The
experimental polymer concentration  is much larger than our
simulation results. This discrepancy can be explained by
considering the depletion force measurements
of~\citet{Wijting2004} in the same colloid-polymer system as was
used in Ref.~\cite{Hoog1999}. They found that the depletion forces
are much smaller than expected, probably due to adsorption of the
polymers on the colloidal surface. We stress that the potentials used in this work compare well with the experiments of~\citet{Ramakrishnan2002} at a size ratio $q$=0.67~\cite{Bolhuis2002}.

Fig.~\ref{fig:bulk}(b) shows the phase diagram in the polymer
reservoir packing fraction $\eta_{p}^{r}$, colloids packing
fraction $\eta_{c}$ representation. The discrepancy between our
results and those of Ref.~\cite{Bolhuis2002} are due to a slightly
different equation of state used for the conversion of the
chemical potential $\mu_{p}$ to the polymer packing fraction in
the reservoir $\eta_{p}^{r}$. In this work, we inverted Eq.
(\ref{EQ:eos1}), while in Ref. \cite{Bolhuis2002} the original SAW
equation of state was used. The binodal has a critical point at
lower $\eta_{p}^{r}$, and the density difference between the gas
and liquid phases increases for increasing $\eta_{p}^{r}$. This
phase diagram is equivalent to the temperature-density phase
diagram of a simple fluid, with the polymer reservoir packing
fraction playing the role of inverse temperature.

\begin{figure}[htbp]
   \centering
   \includegraphics[width=8cm]{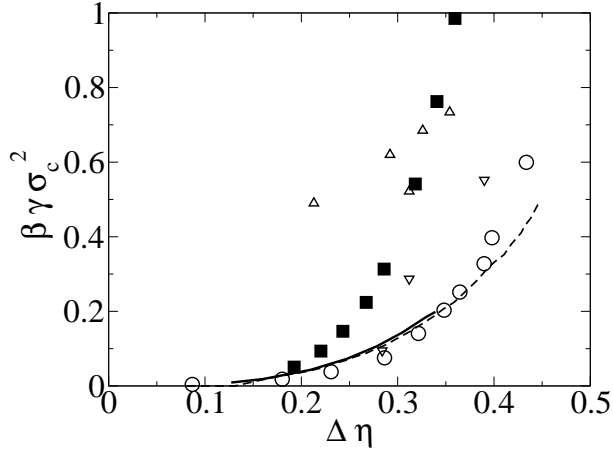}
   \caption{Dimensionless interfacial tension
   $\beta \gamma \sigma_{c}^{2}$ between the gas and liquid phase
   as a function of the difference in packing fractions $\Delta \eta=\eta_{l}-\eta_{g}$ of the
coexisting liquid ($\eta_{l}$)
   and gas ($\eta_{g}$) phase. Results for the interacting polymers with
   size ratio $q=1.05$ (circles) are compared with the results for the AOV
   model with size ratio $q=1.0$ (square). Triangles denote experimental results
   of~\citet{Hoog1999} (triangles up) and~\citet{Aarts2003} (triangles down).
   The thick continuous line indicates the DFT predictions
   of~\citet{Moncho-Jorda2005}, while the dashed lines are the predictions
   of the square gradient approximation theory of~\citet{Aarts2004a}.}
   \label{fig:tensrho}
\end{figure}

In Fig.~\ref{fig:tensrho}, we present the simulation results of
the dimensionless  interfacial tension $\beta \gamma_{gl}
\sigma_{c}^{2}$ for the gas-liquid interface, as a function of the
difference in packing fractions between the gas and the liquid
phase. The interfacial tensions decreases in the case of excluded
volume interactions with respect to the AOV model. The comparison
between our results and the experiments of~\citet{Aarts2003} is
quantitatively better than the results of the AOV model,
although~\citet{Hoog1999} show that it is difficult to obtain
accurate interfacial tension measurements. In addition, we compare
our results with the predictions of the extended free volume
theory plus a square gradient approximation to evaluate the
tension,~\cite{Aarts2004a} and the density functional theory (DFT)
of~\citet{Moncho-Jorda2005}. The DFT uses effective one-component pair potentials  between the colloids that includes the excluded volume interaction according to the approach of~\citet{Louis2002}. The predictions of the
two theories are very close to each other and to our simulation
results.

\begin{figure}[htbp]
   \centering
    \includegraphics[width=7cm]{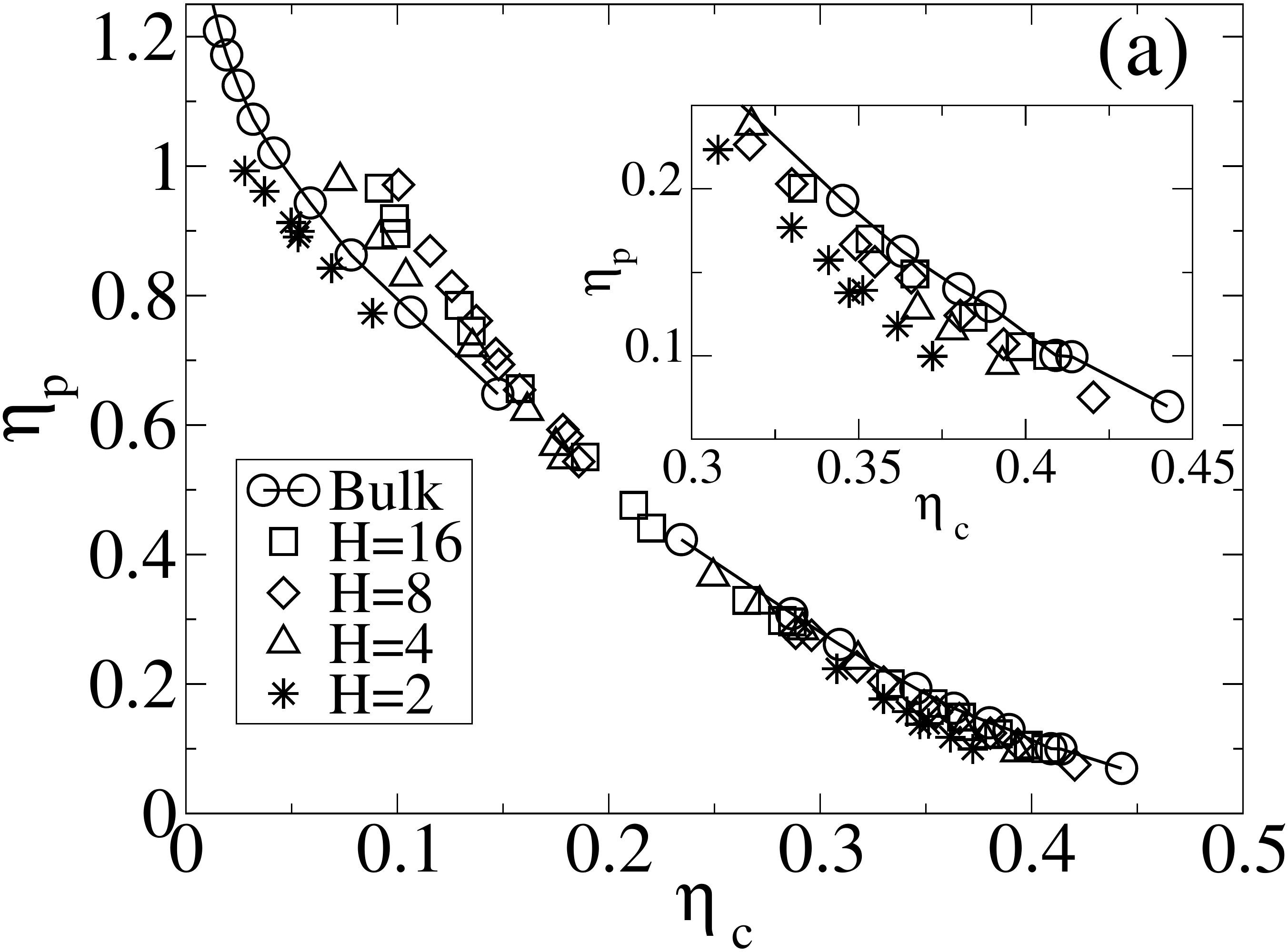}
      \includegraphics[width=7cm]{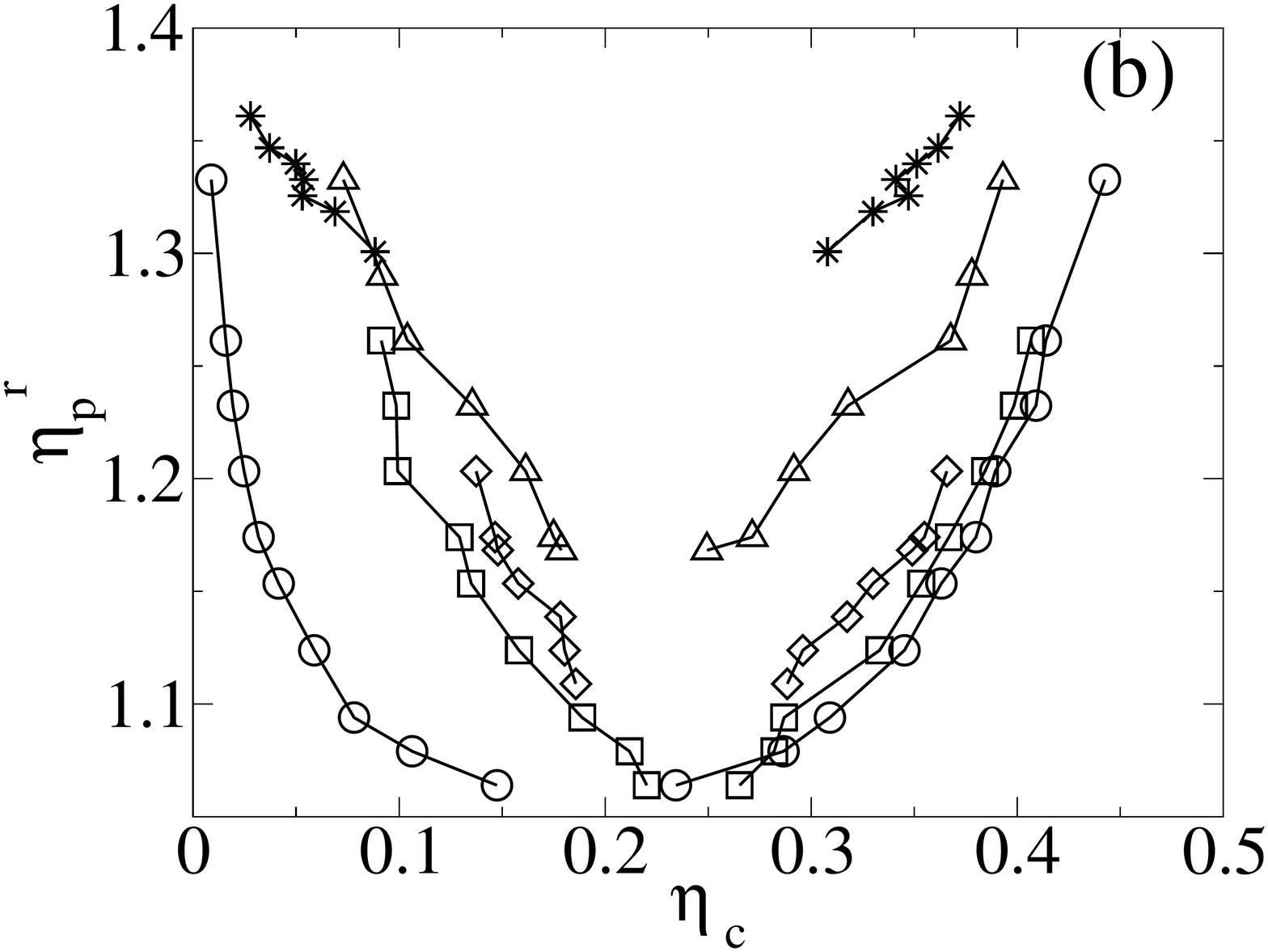}
   \caption{Phase diagram of colloid-polymer mixtures  confined between two hard walls
    with distance $H/\sigma_c=\infty$(bulk), 16, 8, 4,  2. Solid lines are a guide to the eye. 
   (a) Polymer packing fraction $\eta_p$, colloid packing fraction $\eta_c$
   representation. Inset: Blow-up of the high $\eta_c$ region of the binodal. (b) Polymer reservoir packing fraction $\eta_p^r$, colloid
   packing fraction $\eta_c$ representation.  }
   \label{fig:phd1}
\end{figure}

Fig. \ref{fig:phd1} shows the phase diagram of colloid-polymer
mixtures  confined between two hard walls with separation distance
$H/\sigma_c=\infty$, 16, 8, 4, 2. In particular, Fig.~\ref{fig:phd1}(a) shows the phase diagram of colloid-polymer
mixtures in the polymer  packing fraction $\eta_p$, colloid
packing fraction $\eta_c$ representation.
The binodals in Fig.\ref{fig:phd1}(a), hardly change under confinement. We stress that the comparison between the absolute densities inside capillaries of different sizes is complicated due to two factors. First, the density for confined systems is a ill-defined quantity because it depends on the definition of the volume. Our choice of volume depends on the wall separation. Therefore, the comparison between densities for different wall separations in not entirely consistent.
Second, the adsorption of colloids and polymers is different for different wall separations. We find that the colloid adsorption is in general  larger for larger wall separations, and consequently the polymer adsorption is smaller. Therefore, for larger wall separations we expect a larger colloid density and a smaller polymer density. 
The combination of these two effects renders the interpretation of the phase diagram in the $\eta_p$ and $\eta_c$ representation fairly complicated. 
Fig. \ref{fig:phd1}(b)
displays the phase diagram of colloid-polymer mixtures in the
polymer reservoir packing fraction $\eta_p^{r}$, colloid packing
fraction $\eta_c$ representation. The critical points of the
confined systems shift towards higher $\eta_{p}^{r}$ for
decreasing wall separation. 
The interpretation of the binodals in the ($\eta_{p}^{r},\eta_{c}$) representation is more straightforward because only $\eta_{c}$ suffers of the problems described above. 
\begin{figure}[htbp]
   \centering
    \includegraphics[width=4cm]{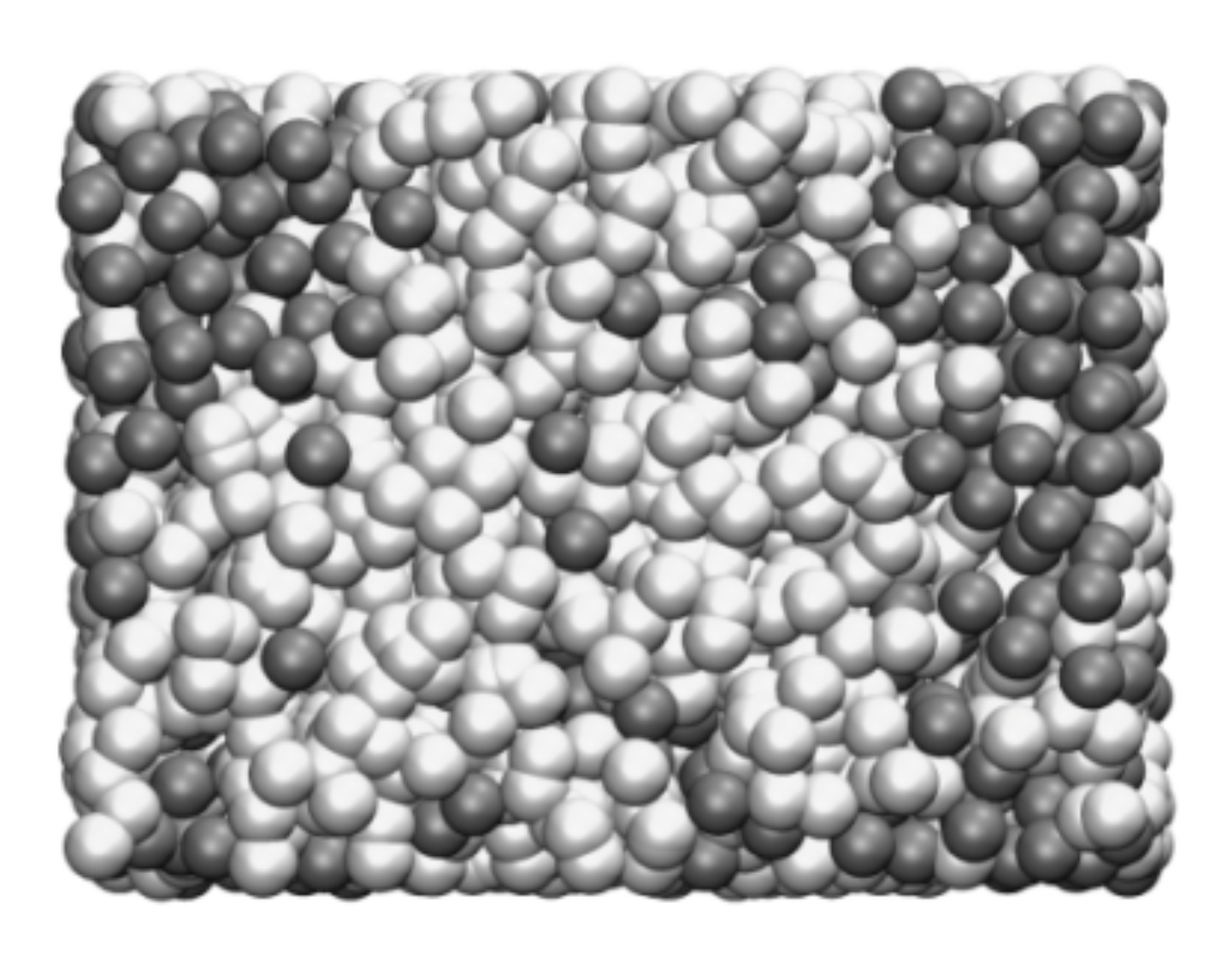}
          \includegraphics[width=4cm]{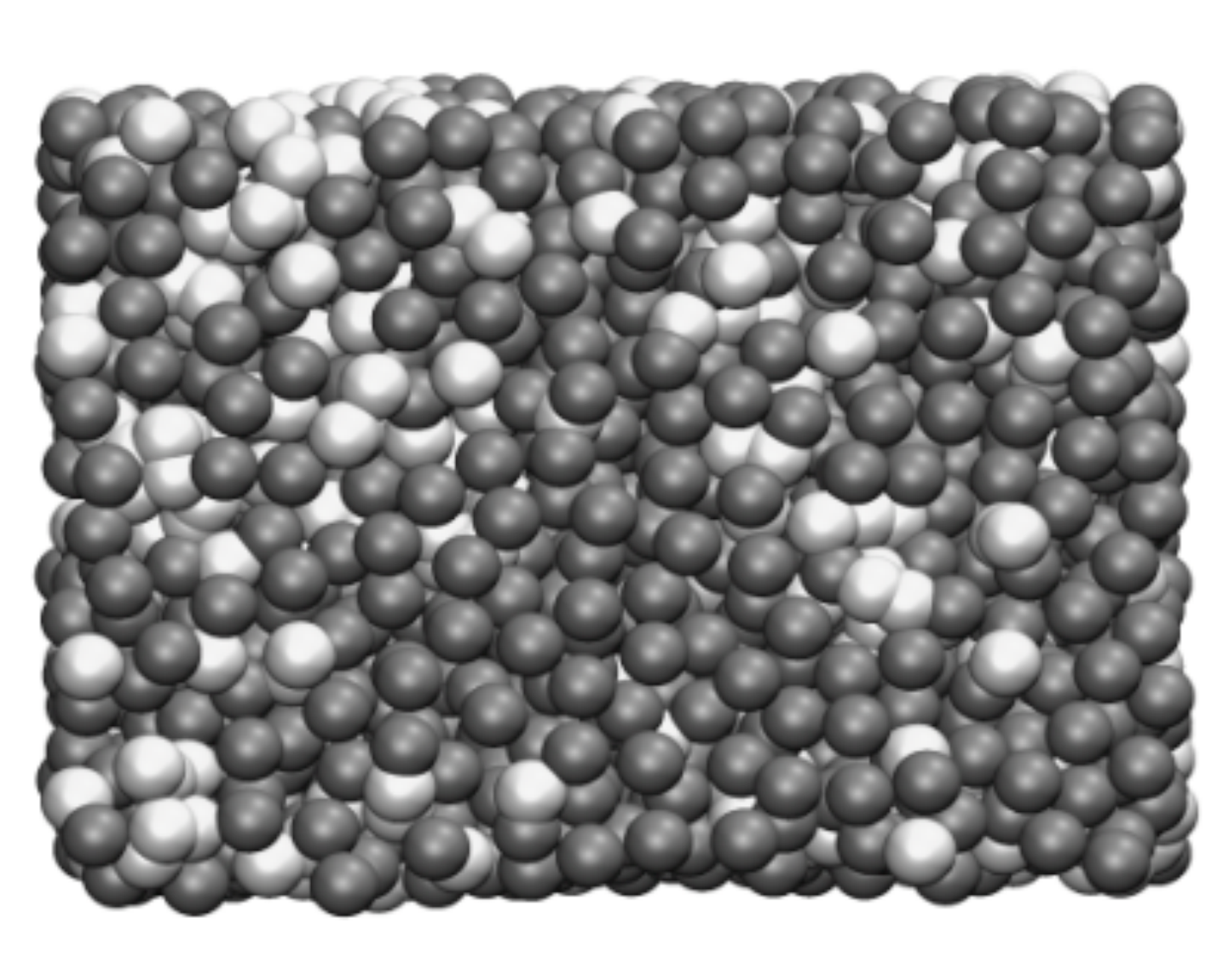}
          \includegraphics[width=4cm]{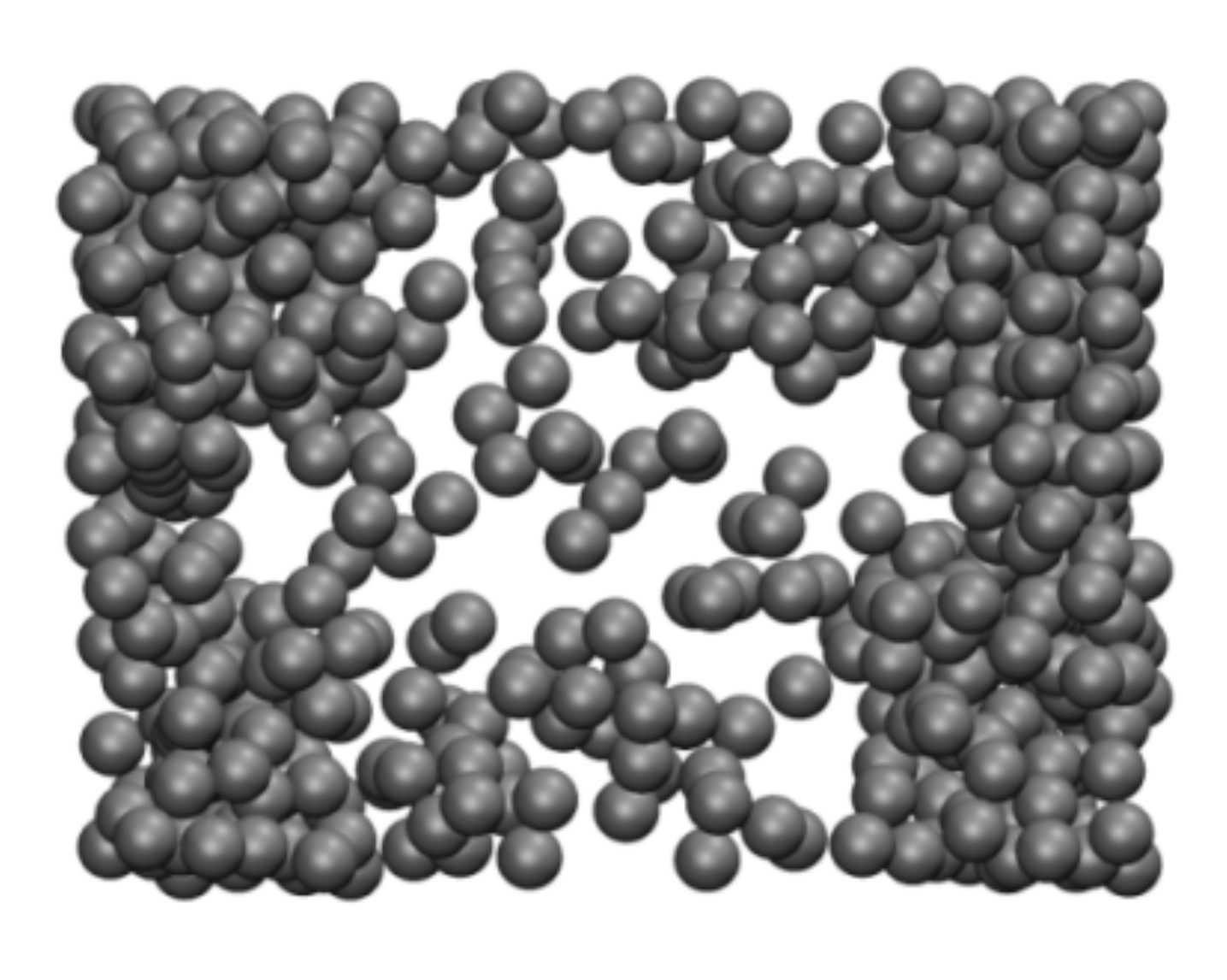}
      \includegraphics[width=4cm]{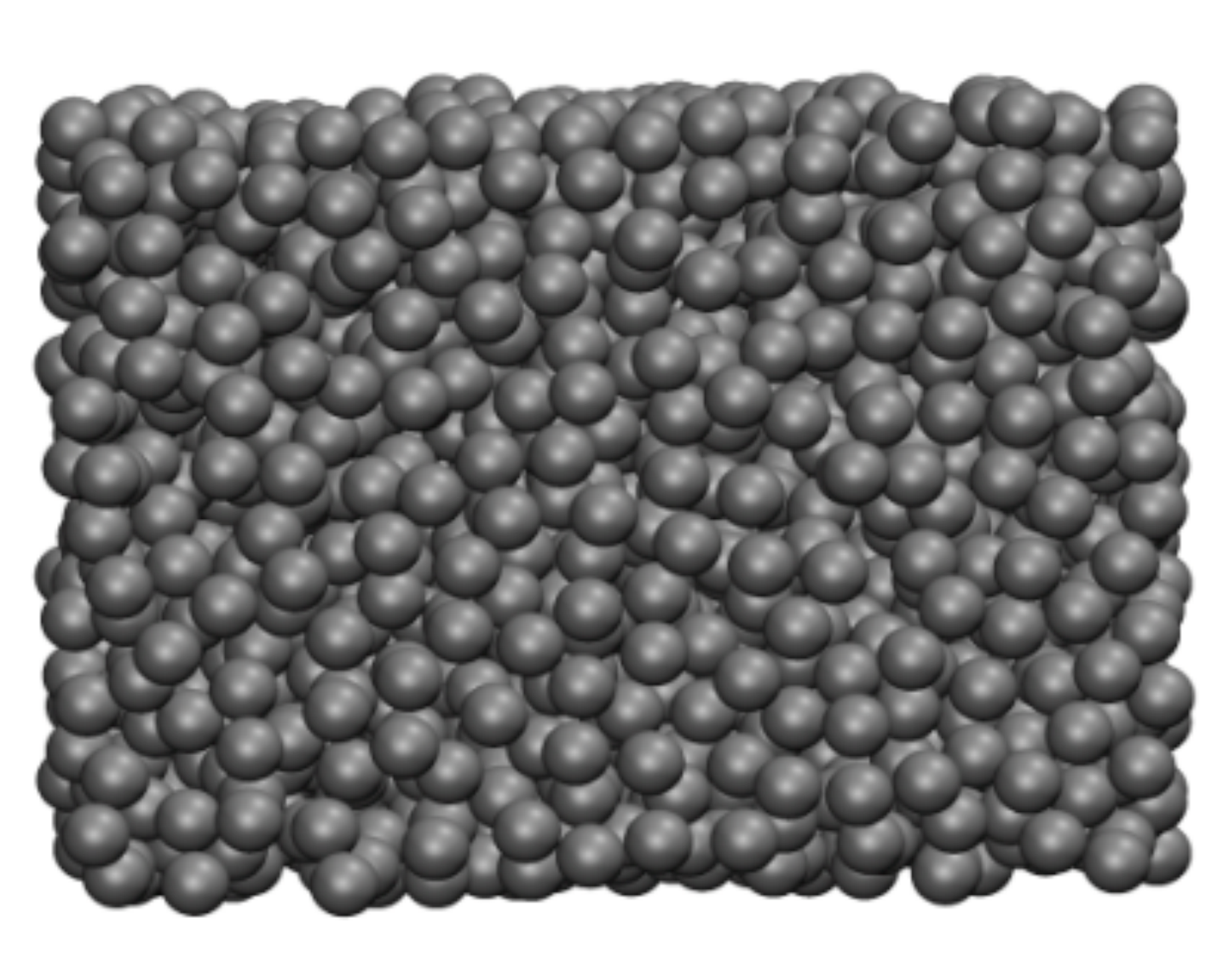}
   \caption{Typical configurations from computer simulations of the coexisting gas (a),(c) and
   liquid phase (b),(d) for the confined system with separation distance $H/\sigma_c =$16
   and chemical potentials $\mu_{c}=10.38$ and $\mu_{p}=5.6$ ($\eta_{p}^{r}$=1.086).
   Colloids are dark grey and polymers are light grey. For clarity
   (c) and (d) display the same configurations of (a) and (b), respectively, without the polymers.}
   \label{fig:snap}
\end{figure}
We can say that the huge shift in gas densities at coexistence for the confined
systems with respect to the bulk binodal is due to the formation
of liquid layers at the walls (See Fig.~\ref{fig:snap}(a) and (c)).
The adsorption of colloids  at the  walls is due to the depletion
attraction, which was also observed in the AOV model. However, the
thickness of the liquid layer was in that case much smaller than
in the present case  and the shift in gas density less
pronounced.~\cite{Schmidt2003,Fortini2006a}

\begin{figure}[htbp]
   \centering
    \includegraphics[width=7cm]{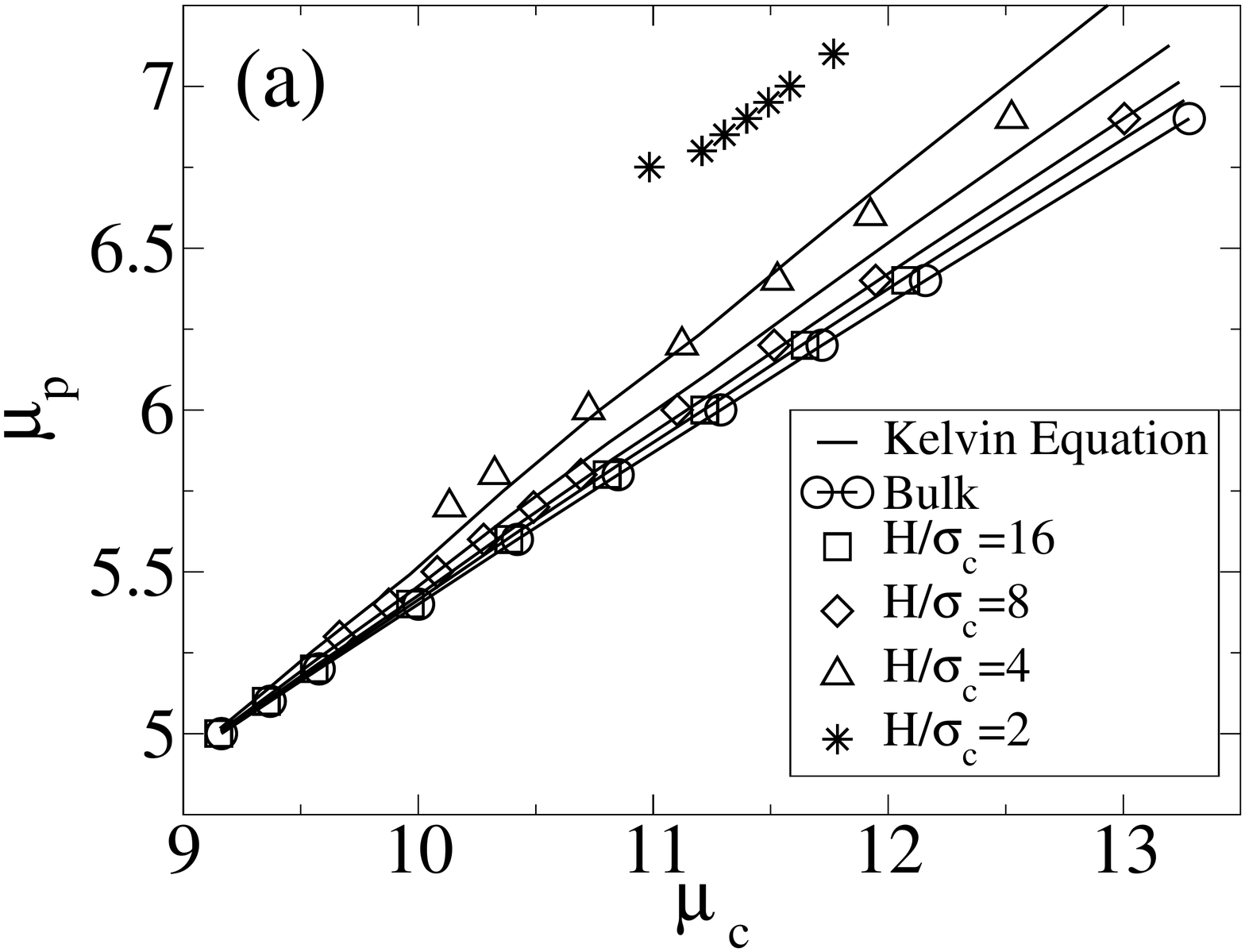}
     \includegraphics[width=7cm]{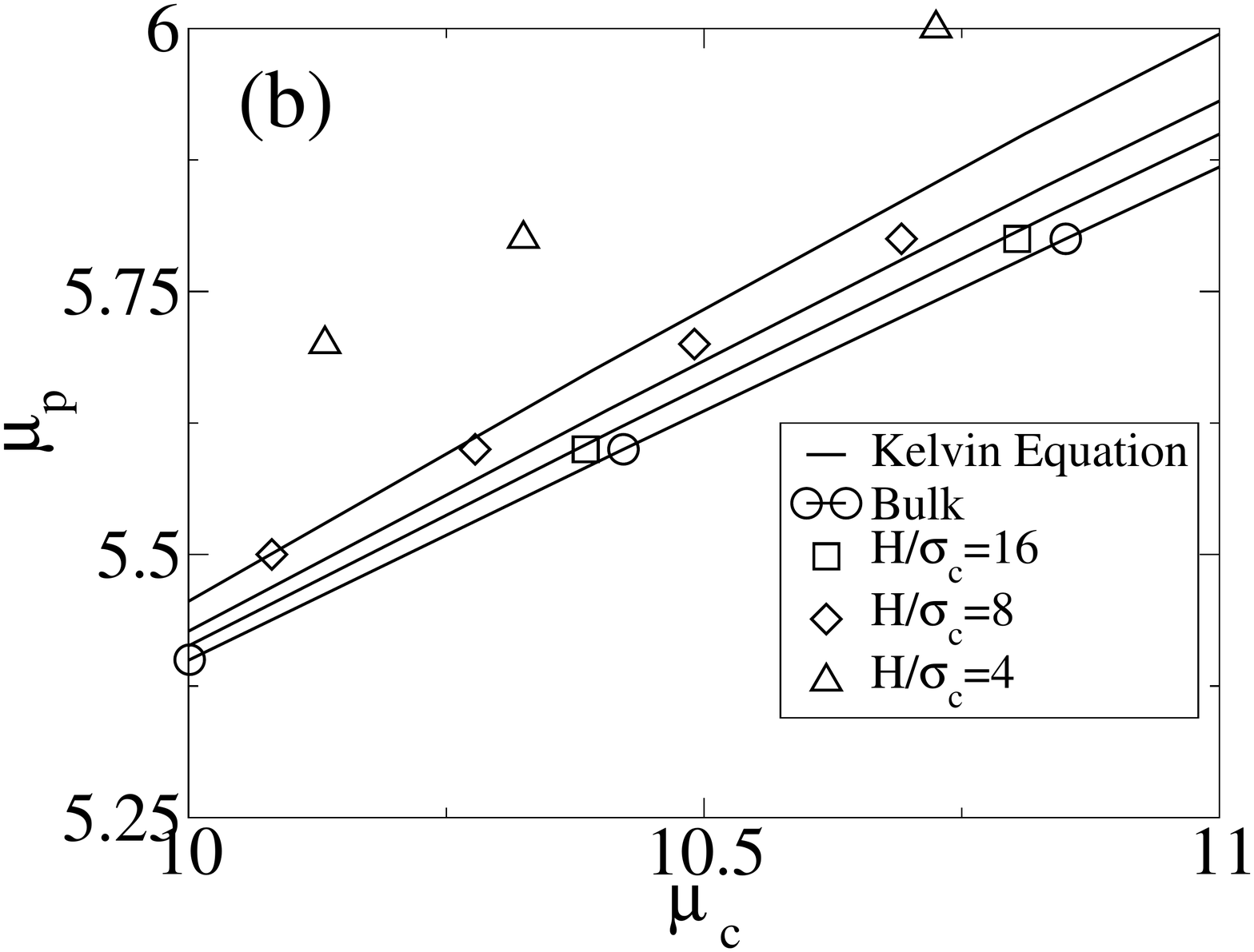}
   \caption{Phase diagram of colloid-polymer mixtures  confined between two hard walls
   with distance $H/\sigma_c=\infty$, 16, 8, 4,  2, in the polymer chemical potential
   $\mu_p$, colloid chemical potential $\mu_c$ representation. In (b) we show a blow-up
   of the phase diagram. For clarity the results of $H/\sigma_c=2$ are not shown.}
   \label{fig:phd2}
\end{figure}

Fig. \ref{fig:phd2} displays the phase diagram in the polymer
chemical potential  $\mu_p$, colloid chemical potential $\mu_c$
representation. These variables do not depend on the definition of the volume and are therefore independent of the wall separation distance. This is the ideal representation to compare the binodals for different wall separation distances. 
The binodals collapse to a single line because of
the thermodynamic equilibrium conditions of the gas-liquid
coexistence. Regions above the binodal are gas-like, while regions
below the binodal are liquid-like. We find a shift of the binodals
towards higher polymer chemical potential and smaller colloid
chemical potential indicating the occurrence of capillary
condensation. Our estimates for the critical points are reported
in Tab.~\ref{tab:crit}.

\begin{table}[h]
\caption{Critical values of the polymer reservoir packing fraction  $\eta^{r}_{p}$,
colloid packing fraction $\eta_{c}$, and chemical potentials $\mu_p$ and $\mu_c$ for
wall separation distances $H/\sigma_{c}=\infty$, 16, 8 , 4, and 2.}
\begin{center}
\begin{tabular}{ c | c c |  c c}
 \hline
$H/\sigma_{c}$ & $(\eta^{r}_{p})_{\rm cr}$ &  $(\eta_{c})_{\rm cr}$&
$(\mu_{p})_{\rm cr}$ & $(\mu_{c})_{\rm cr}$ \\
\hline
 $\infty$ &   1.06(5)  & 0.192(5) &  4.99(5)   &  9.23(5) \\
    16 &    1.08(5)  & 0.243(5) &  5.10(5)   &  9.45(5) \\
      8 &    1.09(5)  & 0.235(5) &  5.16(5)   &  9.50(5)\\
      4 &    1.17(5)  & 0.219(5) &  5.72(5)   &  10.25(5) \\
      2 &    1.27(5)  & 0.199(5) &  6.43(5)   &  10.64(5)\\
\hline
\end{tabular}
\end{center}
\label{tab:crit}
\end{table}

Also shown in Fig. \ref{fig:phd2} are the predictions of the Kelvin equation~\cite{Fortini2006a}
\begin{eqnarray}
  \mu_{\rm c} &=& \mu^{\rm Bulk}_{\rm c} +\frac{2}{h} (\gamma_{\rm wl}-\gamma_{\rm wg})
   \frac{\rho_{\rm c}^{\rm l}-\rho_{\rm c}^{\rm g}}
   {(\rho_{\rm c}^{\rm l}-\rho_{\rm c}^{\rm g})^2  +
     (\rho_{\rm p}^{\rm l}-\rho_{\rm p}^{\rm g})^2}, \nonumber \\
   \mu_{\rm p} &=& \mu^{\rm Bulk}_{\rm p}+ \frac{2}{h} (\gamma_{\rm wl}-\gamma_{\rm wg})
   \frac{\rho_{\rm p}^{\rm l} - \rho_{\rm p}^{\rm g}}
   {(\rho_{\rm c}^{\rm l}-\rho_{\rm c}^{\rm g})^2 +
     (\rho_{\rm p}^{\rm l}-\rho_{\rm p}^{\rm g})^2},
     \label{eq:k}
\end{eqnarray}
The predictions of the Kelvin equation are in good agreement with
the simulation results  for $H/\sigma_c =$16, and 8, but
underestimates the shift for  $H/\sigma_c =$4, and 2.

\begin{figure}[htbp]
   \centering
    \includegraphics[width=7cm]{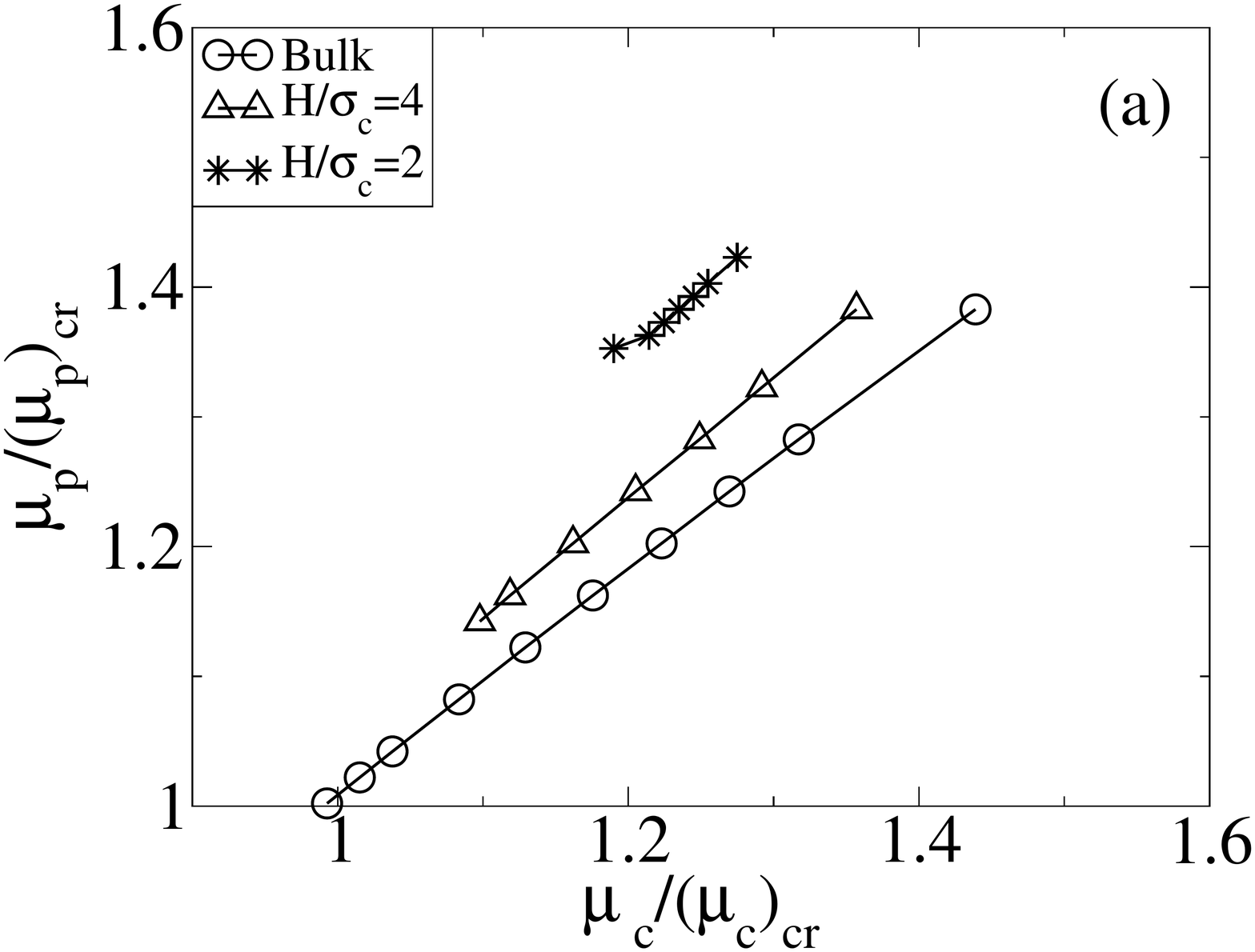}
          \includegraphics[width=7cm]{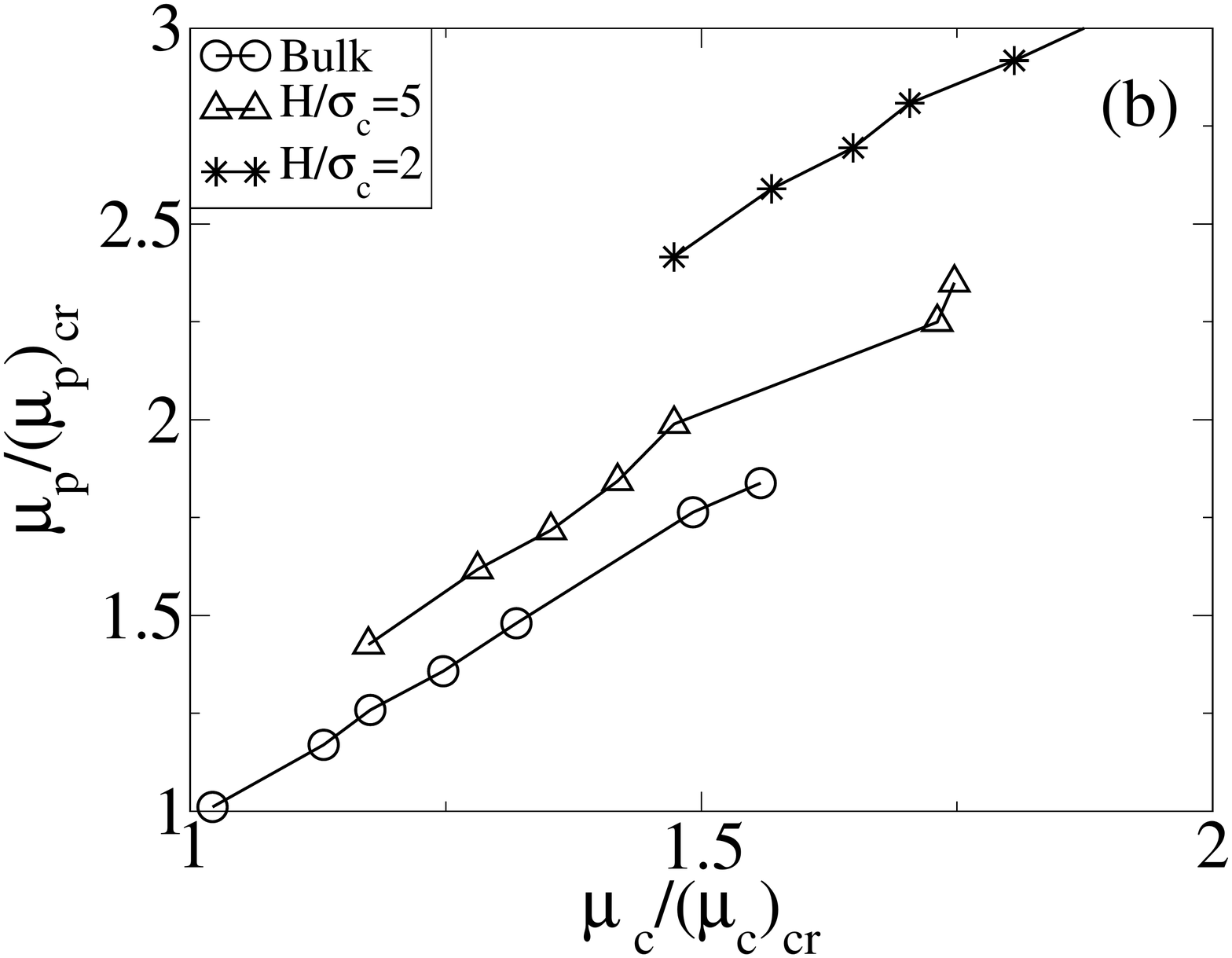}
   \caption{Phase diagram of colloid-polymer mixtures  confined between two hard walls
   in the polymer chemical potential $\mu_p/(\mu_{p})^{\rm bulk}_{\rm cr}$, colloid
   chemical potential $\mu_c/(\mu_{c})^{\rm bulk}_{\rm cr}$ representation. (a)
   $H/\sigma_c =\infty$, 4, and 2 for the model discussed in this paper. (b)
   $H/\sigma_c =\infty$, 5, and 2 for the AOV model. Results are taken from
   Ref.~\cite{Fortini2006a}. }
   \label{fig:comp}
\end{figure}

In order to compare  the results with those for the AOV model we
scale the binodals by the bulk  critical point $(\mu_{p})^{\rm
bulk}_{\rm cr}$ and $(\mu_{c})^{\rm bulk}_{\rm cr}$, reported in
Tab.~\ref{tab:crit}. Fig.~ \ref{fig:comp}(a) and (b) show that the
shift of the binodals and critical points is smaller for the model
with interacting polymers than for the AOV model studied in
Refs.~\cite{Schmidt2003,Fortini2006a} for all state points considered. 
As shown in Fig.~\ref{fig:bulk}, the difference $\rho_{c}^{l}-\rho_{c}^{g}$ between the colloid packing fraction of the liquid and the gas at coexistence is smaller for the interacting polymers than for the AOV model for low $\eta_{p}^{r}$, but larger for high $\eta_{p}^{r}$. On the other hand, Fig.~ \ref{fig:comp}(a) and (b) show that the shift in chemical potential is always smaller for the interacting polymers than for the AOV model. Therefore, we deduce from Eq.~(\ref{eq:k})  that the difference  $\gamma_{\rm wl}-\gamma_{\rm wg}$ (liquid-wall and gas-wall
interfacial tensions at coexistence) is smaller for the interacting polymers model. 

\citet{Moncho-Jorda2005} have shown that at constant
$\eta_{p}^{r}$ the excluded volume interactions increase the
difference in wall tensions. This is not in contrast with our
findings, because for capillary condensation we need to consider
the difference in wall tension at bulk coexistence, that is at
different $\eta_{p}^{r}$ for the AOV model and the interacting
polymer model.

\section{Conclusions}
We have investigated bulk and confined colloid-polymer mixtures,
using Monte Carlo simulations. Colloids are treated as hard
spheres, while polymers were described as soft repulsive  spheres.
Colloid-polymer, polymer-polymer, and wall-polymer
density-dependent interactions were described by the
coarse-grained potentials derived in Ref.~\cite{Bolhuis2002a}. We find
a bulk phase behavior consistent with the findings
of~\citet{Bolhuis2002}. Our results for the bulk phase behavior
are also similar to those for the AOV model with interacting
polymers \cite{Aarts2002}, but here the binodal line lies at
higher polymer packing fractions, i.e., the number of polymers
needed for the demixing transition is larger. These results are in
agreement with the findings of other
authors~\cite{Bolhuis2002,Moncho-Jorda2003,Aarts2004a,Vink2005}.
The comparison of our phase diagram with
experiments~\cite{Hoog1999} is found to be poor for the size ratio
$q=1.05$ considered here. This is surprising, since the same
interaction potentials provided good agreement with experiments~\cite{Ramakrishnan2002} at
a smaller size ratio~\cite{Bolhuis2002}. In fact, this
discrepancy can be explained by considering the results
of~\citet{Wijting2004} on depletion potential measurements on the
same colloid-polymer mixtures that was used in the phase behavior
experiments. These measurements concluded that the depletion
attraction was smaller than expected, probably due to polymer
adsorption on the surface of the colloids.

On the other hand, better agreement is found for the gas-liquid
interfacial tension when compared to the experiments
of~\citet{Aarts2003} for the same system. Our results show that
the gas-liquid interfacial tension is smaller for the interacting
polymers than for the AOV model. This is in agreement with the
works of others on colloid-polymer mixtures with interacting
polymers~\cite{Aarts2004a,Vink2005a,Vink2005,Moncho-Jorda2005}.
Both the square gradient approximation and the DFT provide a good
description of the simulation results.

In addition, we studied the phase behavior of the mixture
confined between two parallel hard walls with separation distance
$H/\sigma_c =$16, 8, 4, and 2. We find that the hard walls induce
capillary condensation, and that the theoretical predictions of
the Kelvin equation  are in reasonable agreement with the
simulation results for $H/\sigma_c =$16 and 8, but underestimate
the binodal shifts for $H/\sigma_c =$4 and 2. Compared to the AOV
model the excluded volume interactions suppress the capillary
condensation. This implies that the effect can only be observed at
statepoints close to bulk coexistence and that smaller plate
separations are needed to induce capillary condensation  for fixed
supersaturation compared with non-interacting polymers.  In other
words the Kelvin length is smaller for interacting polymers than
for the AOV model. In addition, we observed the formation of
rather thick wetting layers at the largest wall separation we
studied. This effect seems to be enhanced by the presence of the
excluded volume interactions. At large wall separations the
wetting layers provide an effective confinement $H-2t$, with $t$
the thickness of the wetting layer, as was recognized
by~\citet{Derjaguin1940}, which increases effectively the Kelvin
length~\cite{Aarts2004b} for the colloid-polymer mixtures with
excluded volume interactions. The influence of excluded volume
interactions on the wetting properties of colloid-polymer mixtures
is currently under investigation.

\acknowledgments We thank Remco Tuinier for discussions and
Arturo Moncho Jord\'a and Dirk G. A. L. Aarts  for their
interfacial tension data. This work is part of the research
program of the {\em Stichting voor Fundamenteel Onderzoek der
Materie} (FOM), that is financially supported by the {\em
Nederlandse Organisatie voor Wetenschappelijk Onderzoek} (NWO). We
thank the Dutch National Computer Facilities foundation for
granting access to the LISA supercomputer. NWO-CW is acknowledge for the TOP-CW funding.

\bibliographystyle{apsrev}
\bibliography{ref_incp}

\end{document}